%% file: main.tex
\documentclass[trackchanges,preprint,twocolumn]{aastex7} % 631 to 7 4/8/2025 COC twocolumn linenumbers 

\newcommand{\objname}{2000 OG$_{44}$} % 9/20/2023 COC
\newcommand{\objnameFull}{(18916) 2000 OG$_{44}$} % 9/20/2023 COC
\usepackage{acronym}

\usepackage{graphicx}
\usepackage[outline]{contour} % 8/7/2022 COC -- outline text
\usepackage[percent]{overpic} % 8/14/2021 COC superimpose text on pictures

\newcommand{\labelcolor}{white} % 8/24/2023 COC; CAT doesn't like yellow 9/14/2023; HHH doesn't like green.; COC tried: white (bad idea for images missing image area); blue, red, and orange too dark; assume magenta will annoy someone; cyan nicer than expected, leaving for now 9/14/2023

\newcommand{\labelpicA}[5]{
\begin{overpic}[width=#4\linewidth]{#1}
	\put (5,7) {\huge\color{\labelcolor} \textbf{\contour{black}{#2}}}
	\put (55,8) {\large\color{\labelcolor} \textbf{\contour{black}{#3}}}
	\put (2,49) {\includegraphics[width=0.15\linewidth]{#5}}
\end{overpic}
}

\begin{document}

\title{A Dormant Captured Oort Cloud Comet Awakens: (18916) 2000 OG44} % 9/19/2023 COC

\correspondingauthor{Colin Orion Chandler}
\email{coc123@uw.edu}

\author[0000-0001-7335-1715]{Colin Orion Chandler}
\email{coc123@uw.edu}
\affiliation{Dept. of Astronomy \& the DiRAC Institute, University of Washington, 3910 15th Ave NE, Seattle, WA 98195, USA}
\affiliation{\acs{LSST} Interdisciplinary Network for Collaboration and Computing, 933 N. Cherry Avenue, Tucson, AZ 85721, USA}
\affiliation{Dept. of Astronomy \& Planetary Science, Northern Arizona University, PO Box 6010, Flagstaff, AZ 86011, USA}

\author[0000-0001-5750-4953]{William J. Oldroyd}
\email{william.oldroyd@nau.edu}
\affiliation{Dept. of Astronomy \& Planetary Science, Northern Arizona University, PO Box 6010, Flagstaff, AZ 86011, USA}

\author[0000-0001-9859-0894]{Chadwick A. Trujillo}
\email{chad.trujillo@nau.edu}
\affiliation{Dept. of Astronomy \& Planetary Science, Northern Arizona University, PO Box 6010, Flagstaff, AZ 86011, USA}

\author[0009-0007-1972-5975]{Dmitrii E. Vavilov}
\email{vavilov@uw.edu}
\affiliation{Dept. of Astronomy \& the DiRAC Institute, University of Washington, 3910 15th Ave NE, Seattle, WA 98195, USA}

% WAB contributions: analysis, methods section writing
\author[0000-0002-6023-7291]{William A. Burris}
\email{wab77@nau.edu}
\affiliation{Dept. of Astronomy \& Planetary Science, Northern Arizona University, PO Box 6010, Flagstaff, AZ 86011, USA}

\begin{abstract}
We report the discovery of activity emanating from \objnameFull{} (alternately designated 1977 SD), a minor planet previously reported to be both an extinct comet or an \acl{ACO}. We observed \objname{} with a thin tail oriented towards the coincident anti-solar and anti-motion vectors (as projected on the sky) in images we acquired on UT 2023 July 24 and 26 with the \acl{APO} 3.5~m \acl{ARC} telescope (New Mexico, USA). We also include observations made in Arizona with the \acl{VATT} at the \acl{MGIO} and the Lowell Observatory \acl{LDT} near Happy Jack. We performed dynamical simulations that reveal \objname{} most likely originated in the Oort cloud, arriving within the last 4 Myr. We find \objname{}, which crosses the orbits of both Jupiter and Mars, is at present on an orbit consistent with a \acl{JFC}. We carried out thermodynamical modeling that informed our broader diagnosis that the observed activity is most likely due to volatile sublimation. 
\end{abstract}

\keywords{
Asteroids (72), % keeping asteroid belt as the object is listed as an ACO 10/27/2023 -- also these are vestigial anyhow
Comae (271), 
Comet tails (274),
Hilda group (741),
Short-period comets (1452),
Comet dynamics (2213)
}

\acresetall % our new friend; resets memory of used acronyms for TeX package 11/1/2023 COC

%%%%%%%%%%%%%%%%%%%%%%%%%%%%%%%%%%%%%%%%%%%%%%%%%%%%%%%%%%%%%%%%
\section{Introduction} \label{sec:intro}

Comets are not expected to exhibit tails or comae at all times. Most commonly, comets will return to a quiescent state when too distant from the Sun for sufficient energy to induce sublimation for the involved ice species. These same objects may reactivate when enough energy is available, when they are closer to the Sun. Some objects on orbits consistent with comet-class objects, including short-period comets like the \acp{JFC}, have never been observed to be active, despite extensive observational efforts. Some of these objects, 
% These objects may be referred to as 
the asteroids on cometary orbits (\acs{ACO}; \citealt{fernandezAlbedosAsteroidsCometLike2005,licandroMultiwavelengthSpectralStudy2006}), 
are thought to be asteroids that were perturbed onto cometary orbits, while others, such as 
dormant comets \citep{yeDormantCometsNearEarth2016}, 
extinct comets \citep{fernandezLowAlbedosExtinct2001},  
dead comets or cometary nuclei \citep{lamySizesShapesAlbedos2004}, 
dark comets \citep{seligmanDarkCometsUnexpectedly2023,taylorSeasonallyVaryingOutgassing2024}, 
or 
Manx comets \citep{meech2013P2Pan2014}, 
are thought to have started out as comets but are no longer active. Inactivity may be permanent, for example resulting from volatile depletion, %a process thought to happen to all comets over time, 
or temporary, such as insulation and suffocation by dust or other materials that cover volatiles, thereby quenching the process by reducing available energy \citep{kuhrtFormationCometarySurface1994}.

Some inactive bodies have later been found to display activity. Late activity detection may be the result of any combination of improved observing geometry (e.g., geocentric and heliocentric distance), enhanced observing circumstances (e.g., larger aperture telescope), or onset of activity (due to, for example, perihelion passage), such as (551023) 2012 UQ$_{192}$ \citep{chandlerActiveAsteroidsCitizen2024} and 2009 DQ$_{118}$ \citep{oldroydRecurringActivityDiscovered2023}, a \ac{JFC} and \ac{QHC}, respectively. Quasi-Hildas are minor planets with orbits similar to (153) Hilda, but not themselves in interior 3:2 \ac{MMR} with Jupiter, as is the case for the Hilda asteroids \citep{kresakDynamicalInterrelationsComets1979}.

Both the \acp{JFC} and quasi-Hildas, which are thought to originate primarily in the Kuiper belt \citep{volkScatteredDiskSource2008,emelyanenkoModelCommonOrigin2013}, are dynamically unstable populations with lifetimes on the order of tens of thousands of years %{$10^7$ yr} 
\citep{levisonKuiperBeltJupiterFamily1997}. To date, roughly 700 \acp{JFC} and 20 active quasi-Hildas have been identified. Study of volatile-bearing minor planets helps us understand the origins of terrestrial water \citep[e.g.,][]{morbidelliSourceRegionsTime2000,hsiehPopulationCometsMain2006,hartoghOceanlikeWaterJupiterfamily2011,chandlerChasingTailsActive2022a} and informs planetary defense \citep{levisonKuiperBeltJupiterFamily1997}.

It is thought that only a handful of Oort cloud comets could have been captured \citep{hillsCometShowersSteadystate1981}, in part because most migrating inner Oort cloud ($a<20,000$~au) comets would be deflected by Jupiter (or Saturn) before they are observable, a phenomenon sometimes referred to as the ``Jupiter barrier'' \citep{levisonScatteredDiskSource2006}. There are only a handful of short-period comets thought to originate in the Oort cloud. 96P/Macholz 1 has been identified as both having a highly unusual spectral signature \citep{langland-shulaUnusualSpectrumComet2007,schleicherExtremelyAnomalousMolecular2008} and a notably ``blue'' color by \ac{JFC} standards (e.g., $g'-i'\approx0.5$) that is more similar to the Halley-type comets than \acp{JFC} \citep{eisnerPropertiesBareNucleus2019}; 96P is also highly inclined ($i\approx58^\circ$).

As part of our ongoing campaign to discover activity on objects previously thought to be asteroidal through our NASA Partner Citizen Science program \textit{Active Asteroids}\footnote{\url{http://activeasteroids.net}} \citep{chandlerActiveAsteroidsCitizen2024}, we periodically observe objects reported to be inactive comets. One such object, \objnameFull{} (alternative designation 1977 SD), % CAT: why the 1977 year if it was found in 2000? describe
has been the subject of investigations for decades following its discovery on UT 2000 July 30 by the \ac{LINEAR} survey at Socorro, New Mexico, with precovery observations found dating as early as 1954 \citep{hahnMPEC2000S6520002000}. 

%%%%%%%%%%%%%%%%%%%%%%%%%%%%%%%%%%%%%%%%%%%%%%%%%%%%%%%%%%%%%%%%
\section{Observations}
\label{sec:observations}

\newcommand{\figsize}{0.31}
\begin{figure*}
    \centering
    \begin{tabular}{ccc} 
    \labelpicA{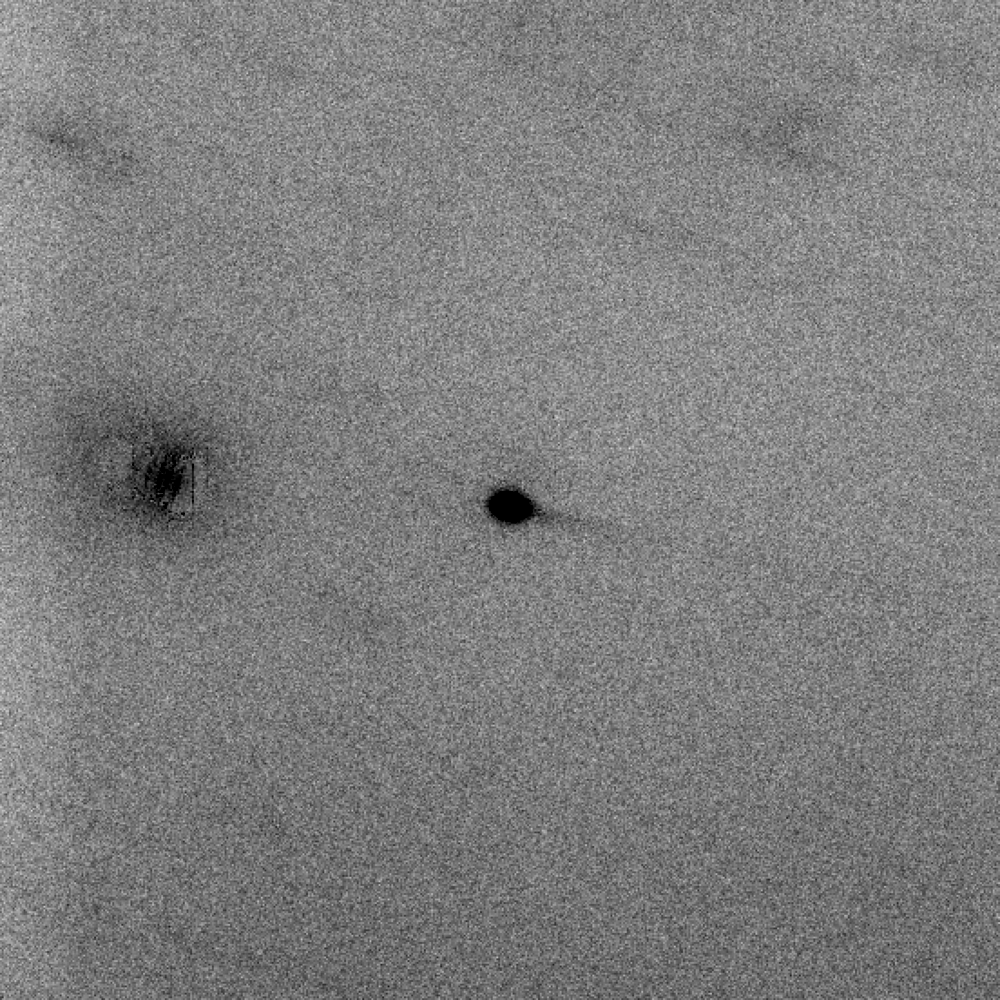}{a}{2023-07-24}{\figsize}{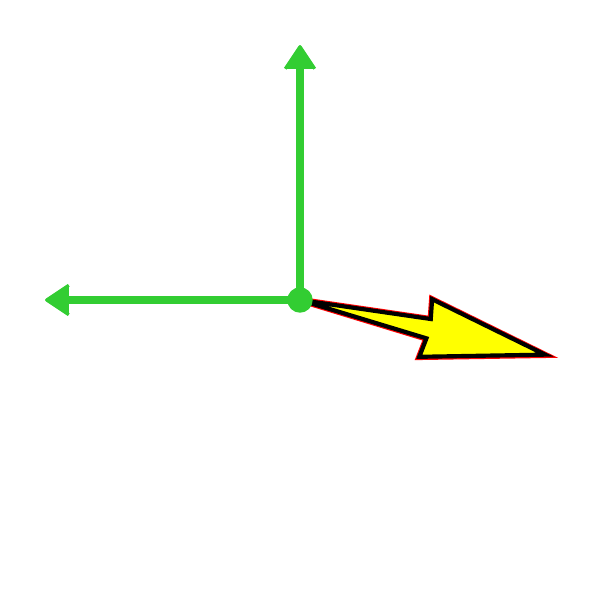} & 
    \labelpicA{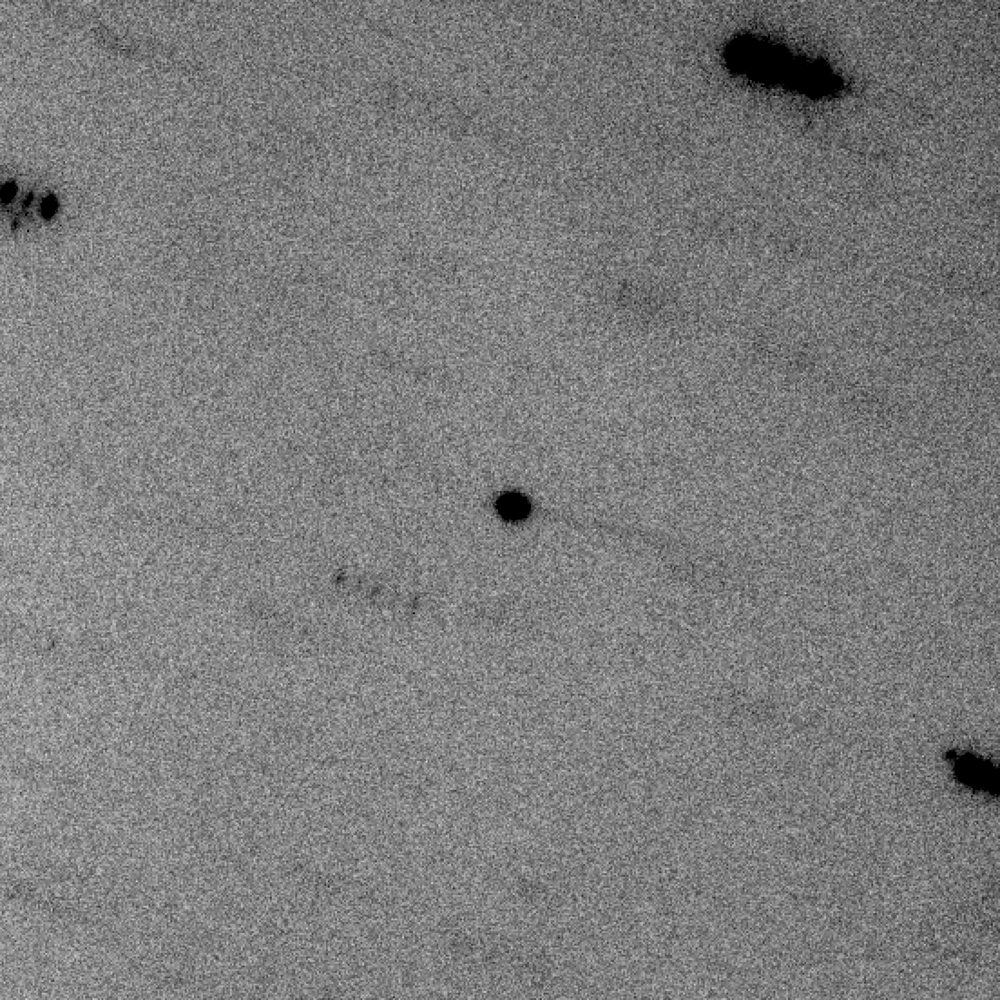}{b}{2023-07-26}{\figsize}{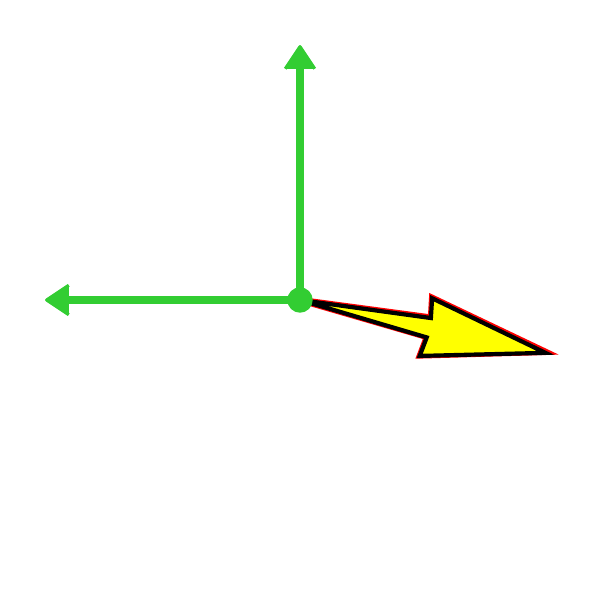} &
    \labelpicA{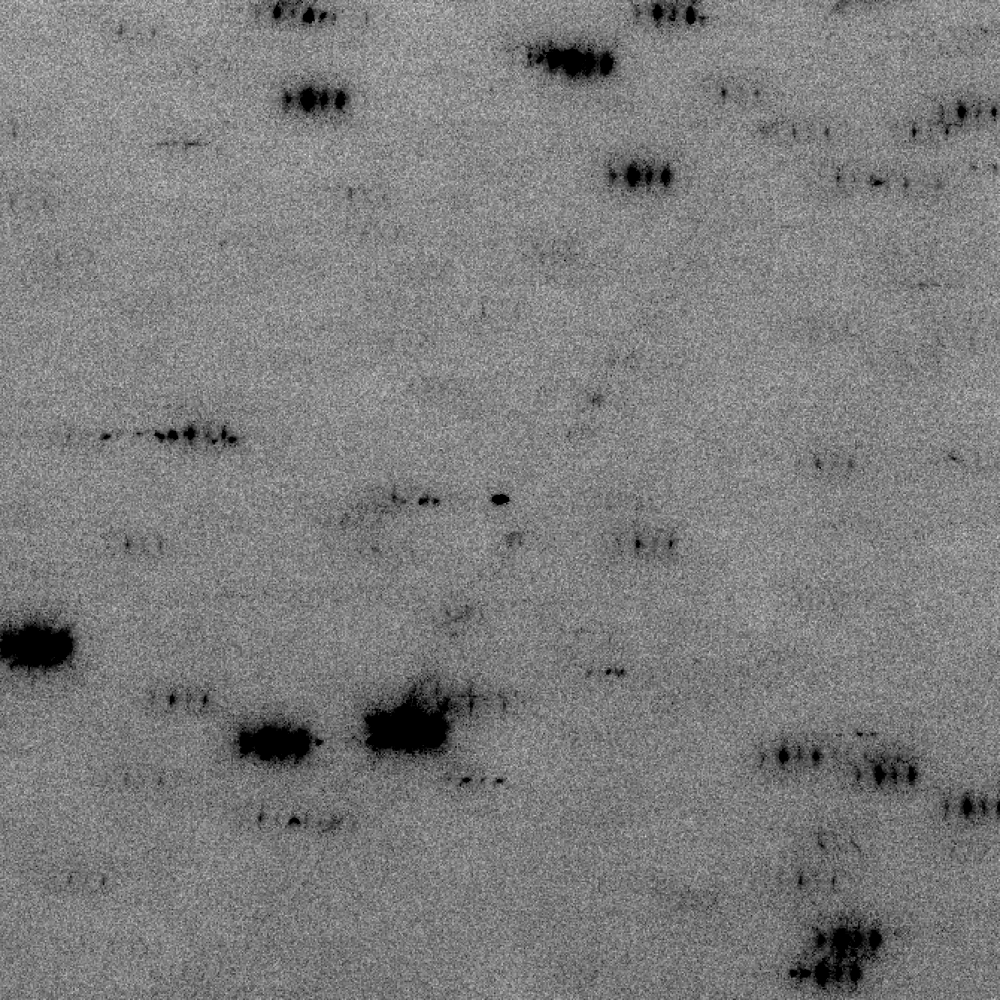}{c}{2023-09-16}{\figsize}{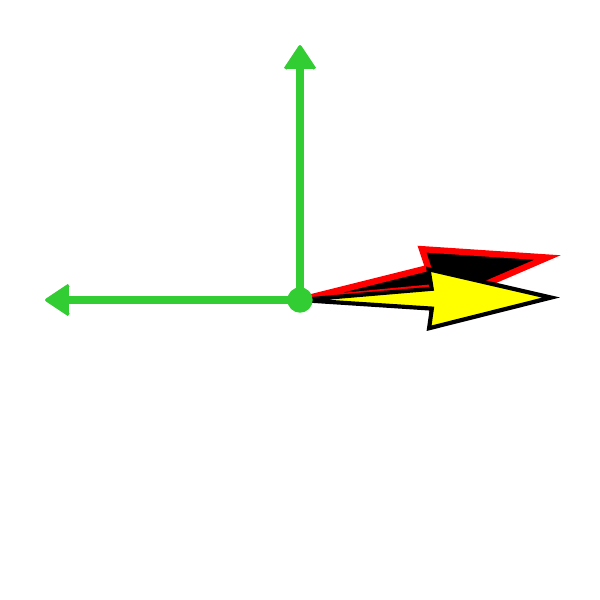}
    \\
    \labelpicA{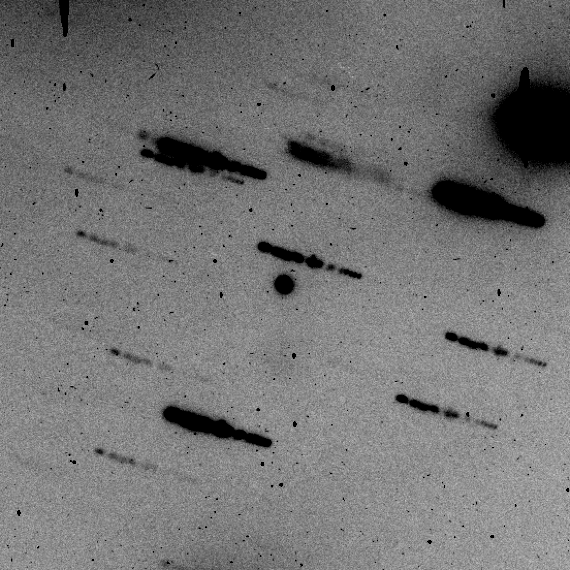}{d}{2023-12-12}{\figsize}{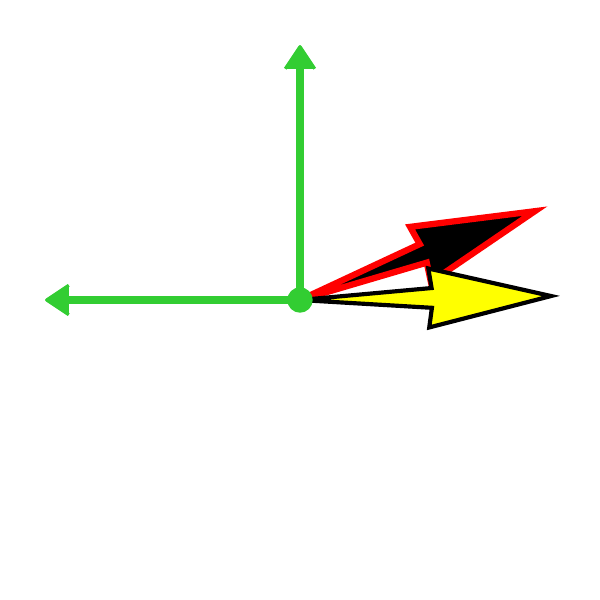} & 
    \labelpicA{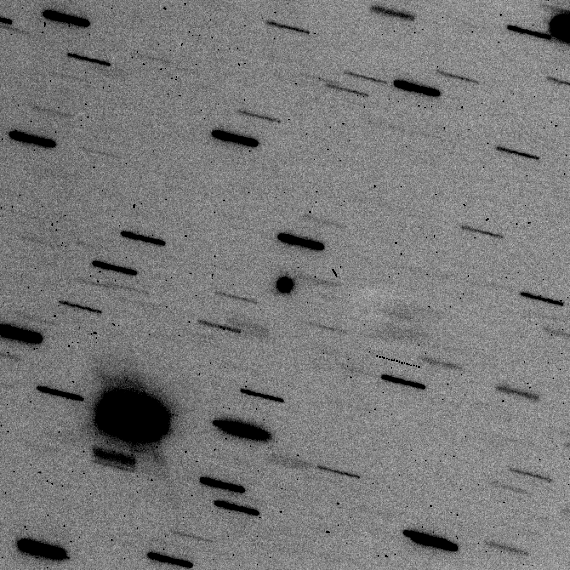}{e}{2023-12-15}{\figsize}{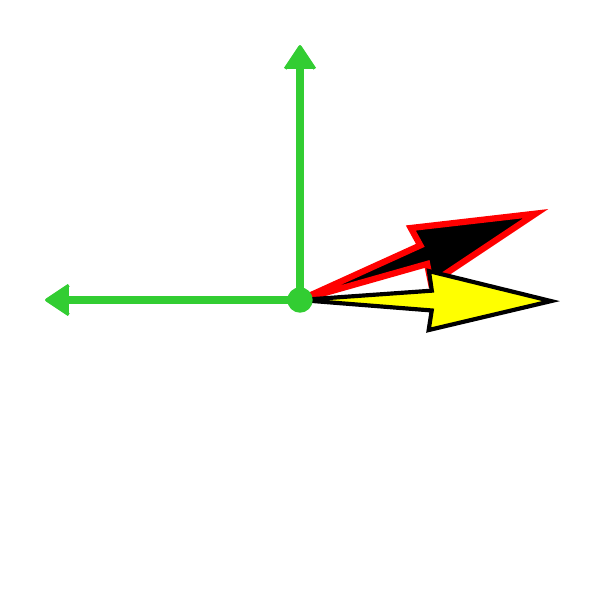} & 
    \labelpicA{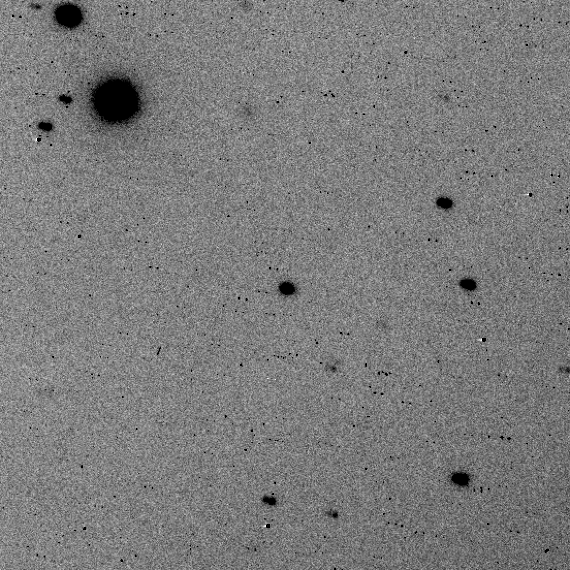}{f}{2023-12-19}{\figsize}{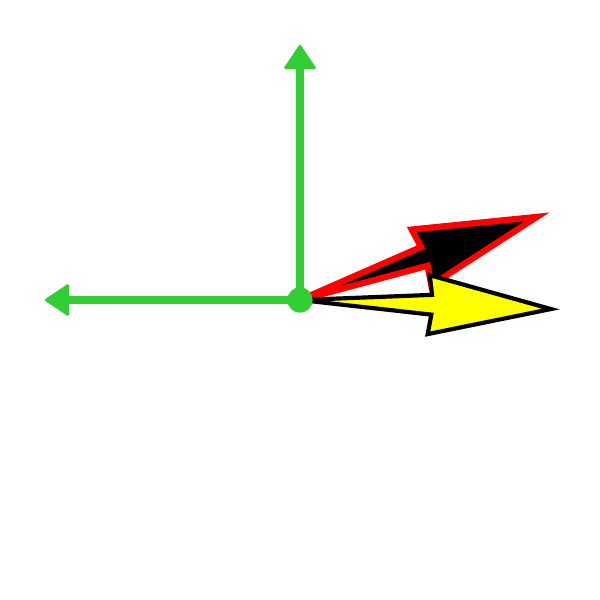}\\
    \labelpicA{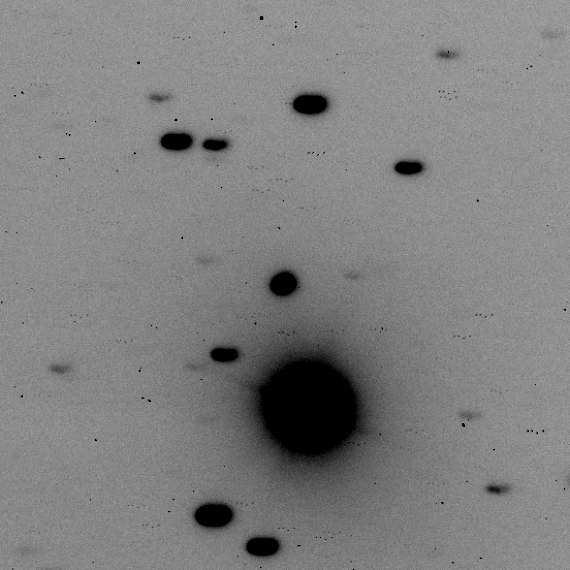}{g}{2024-01-04}{\figsize}{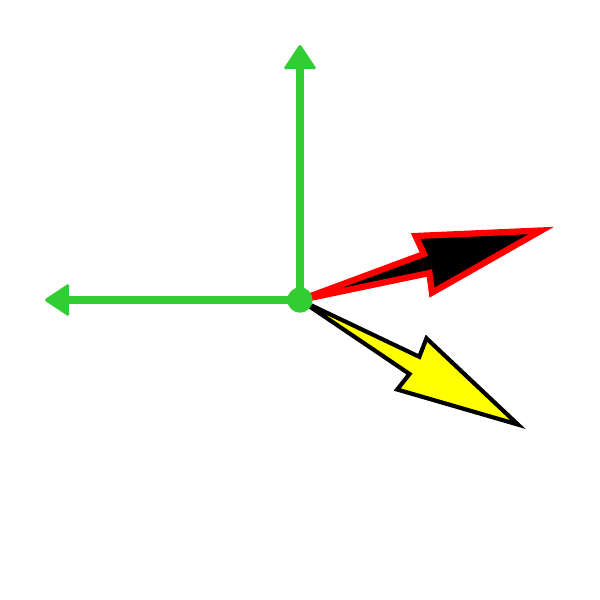} & 
    \labelpicA{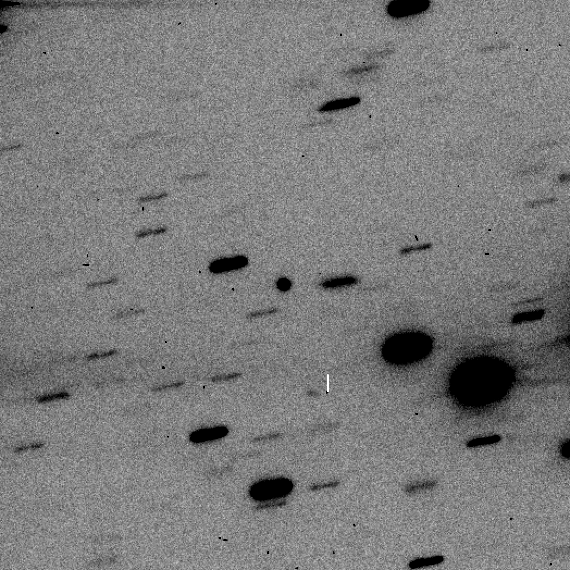}{h}{2024-04-11}{\figsize}{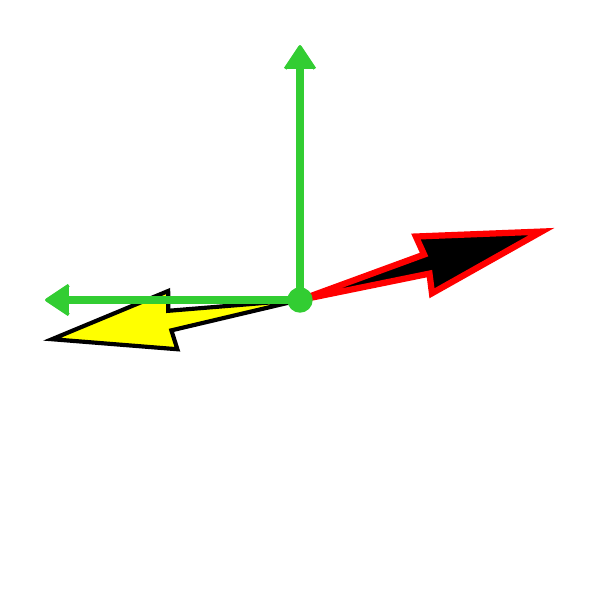}\\
    \end{tabular}
    \caption{\objnameFull\ is centered in these 126\arcsec{}$\times$126\arcsec{} images, with North up and East left. 
    A tail is seen oriented towards the coincident anti-solar (yellow arrow) and anti-motion (red outlined black arrow) directions as projected on the sky in panels \textbf{a} and \textbf{b}, both comprised of \textit{VR}-band images acquired with the \acf{APO} \acf{ARCTIC} instrument on the 3.5~m \acf{ARC} telescope in Sunspot, New Mexico. Panels \textbf{c} -- \textbf{h} do not show evidence of activity, likely due to the cessation of activity on \objname{}. 
    %\added{All \ac{ARCTIC} images acquired under program ID UW06, PI Chandler, observers C. Chandler, W. Oldroyd.}{}
    \textbf{(a)} UT 2023 July 24  $\sigma$-clipped co-addition of 18 \ac{ARCTIC} exposures (3$\times$90~s \textit{VR}-band, plus $2\times 60$~s + $3\times 45~s$ + $5\times 30~s$ $r$-band). % studentized
%\textit{VR}-band exposures with the \ac{APO} \ac{ARCTIC} instrument .
    \textbf{(b)} UT 2023 July 27 co-addition of 6 $\times$ 90~s \ac{ARCTIC} images. 
    \textbf{(c)} UT 2023 September 16, co-added 2 $\times$ 120~s $V$-band plus 2$\times$120~s $R$-band VATT4K images acquired with the 1.8~m \acf{VATT} at \acf{MGIO} in Arizona. %\added{}; \texttt{Siril} ``Studentized'' stacking.}{} % NTS: not reduced yet here
    \textbf{(d)} UT 2023 December 12, $7\times 300$~s plus $5\times 120$~s co-added \textit{VR}-band \ac{ARCTIC} exposures. %; no apparent activity was present. 
    \textbf{(e)} UT 2023 December 15, co-added 18 $\times$ 120~s $V$-band \ac{VATT}4K exposures.
    \textbf{(f)} UT 2023 December 19, co-added 2 $\times$ 150~s \textit{VR}-band \ac{ARCTIC} exposures. 
    \textbf{(g)} UT 2024 January 4, co-added 4 $\times$ 60~s \textit{VR}-band \ac{ARCTIC} images. 
    \textbf{(h)} UT 2024 April 11, co-added 2 $\times$ 300~s $r$-band exposures with \ac{LMI} on the 4.3~m \ac{LDT} near Happy Jack, Arizona.
    }
    \label{fig:activity}
\end{figure*}

\begin{table*}
\centering
\caption{Table of Observations}
\label{tab:observations}
\begin{tabular}{cccrlllcrc}
Active & UT Date       & Site & $t_\mathrm{tot}$ [s] & $N_\mathrm{exp}$ & $t_\mathrm{exp}$ [s]        & Filter     & $r_\mathrm{H}$ [au]    & $\nu$ [$^\circ$]   & Observers \\
\hline
\ *      & 2023-07-24 & 705  & 1125   & 3, 5, 2, 3, 5           & 90, 90, 60, 45, 30 & \textit{VR}, $r$, $r$, $r$, $r$ & 1.649 & 338 & COC, WJO   \\
\ *      & 2023-07-26 & 705  & 540    & 6                   & 90             & \textit{VR}         & 1.645 & 339 & COC, WJO   \\
       & 2023-09-16 & 290  & 480    & 2,2                 & 120,120        & $V$, $R$        & 1.613 & 10  & COC, WJO   \\
       & 2023-12-12 & 705  & 2700   & 5,7                 & 120,300        & \textit{VR}         & 1.915 & 56  & COC, WJO   \\
       & 2023-12-15 & 290  & 2160   & 18                  & 120            & $V$          & 1.931 & 57  & CAT, WAB   \\
       & 2023-12-19 & 705  & 300    & 2                   & 150            & \textit{VR}         & 1.952 & 59  & COC, WJO   \\
       & 2024-01-04 & 705  & 240    & 4                   & 60             & \textit{VR}         & 2.041 & 65  & COC, WJO   \\
       & 2024-04-11 & G37  & 600    & 2                   & 300            & $r$          & 2.644 & 94  & WJO      
\end{tabular}

\raggedright
\footnotesize

* indicates activity was identified in images. $t_\mathrm{tot}$ is the total combined exposure time. 
$N_\mathrm{exp}$ is the number of exposures. 
$t_\mathrm{exp}$ is the exposure time. 
$r_\mathrm{H}$ is the heliocentric distance and $\nu$ the true anomaly angle. 
Site 705 is the \acf{APO} in Sunspot, New Mexico, USA. Site 290 is \acf{MGIO} near Safford, Arizona, USA. Site G37 is the \acf{LDT} near Happy Jack, Arizona, USA.
Observers were C. O. Chandler (COC), C. A. Trujillo (CAT), W. J. Oldroyd (WJO), and W. A. Burris (WAB). 
\end{table*}

We observed \objname{} on eight separate nights with three different telescopes, acquiring 66 images in total (Table \ref{tab:observations}). 
A long thin tail oriented towards the coincident anti-solar and anti-motion vectors (as projected on the sky) emanating from \objname{} is clearly visible in the \ac{ARCTIC} images acquired UT 2023 July 24 and 26 (Figure \ref{fig:activity} panels a and b). At the time these images were acquired, \objname{} was inbound with a heliocentric distance of $r_H=1.649$~au and $r_H=1.645$~au, at true anomaly angles $\nu=338^\circ$ and $\nu=339^\circ$. 

During the time the \ac{VATT} images we acquired (UT 2023 September 16), \objname{} was at $r_H=1.613$~au and $\nu=10^\circ$. We were unable to identify any definitive indicators of activity in those images, possibly due to the combination of the smaller aperture and observing conditions, though our subsequent images with the 3.5~m \ac{APO} 3.5~m and \ac{LDT} 4.3~m telescopes also did not reveal cometary features, suggesting activity had already subsided.

%%%%%%%%%%%%%%%%%%%%%%%%%%%%%%%%%%%%%%%%%%%%%%%%%%%%%%%%%%%%%%%%%%%%%%%%%%%%%%%%%%%%%%%%%%%%%%%%
\section{Orbital and Physical Properties}
\label{sec:properties}

% TODO entire section to table per CAT, WJO 5/18/2024 COC added this comment
As mentioned in Section \ref{sec:intro}, \objname{} has been studied extensively. Here we summarize previous measurements, including those needed for analyses in subsequent sections. 
% \objnameFull{} has a semi-major axis $a=3.858$~au, orbital eccentricity $e=0.584$, inclination $i=7.396^\circ$, perihelion distance $q=1.604$~au with the most recent passage occurring on 2023 August 30, aphelion distance $Q=6.112$~au, a Tisserand Parameter with respect to Jupiter of $T_\mathrm{J}=2.735$ (see Section \ref{sec:class}), and an orbital period of $P=7.558$~yr; 
Orbital parameters (Table \ref{tab:orbitparameters}) and measured properties (Table \ref{tab:params}) are provided in the Appendix. 
% JPL SBDB $H=14.62$ (E2023-TC0). 
At the time \objname{} was discovered (UT 2000 July 30), it was at a heliocentric distance $r_H=2.109$~au, inbound to perihelion with a $\nu=292^\circ$ true anomaly angle.

% The absolute magnitude, albedo, and diameter of \objname{} have been reported many times now. Figure out a clever way of pulling it all in here and citing everyone too.

% Comets II (Lamy) stuff: (possible dead comet): $T_\mathrm{J}=2.74$, $r_n=3.867^{+0.50}_{-0.40}$, $p_v=0.038^{+0.018}_{-0.017}$ oh this is all from \cite{fernandezLowAlbedosExtinct2001}; Lamy refers to OG as an ecliptic comet. Actually, none of this is particularly new I guess, and noting that Fernandez is an author of the book chapter too.

One parameter especially significant for our calculations and activity assessment is the rotation period of \objname{}. \cite{szaboRotationalPropertiesHilda2020} acquired 670 frames of \objname{} with the Kepler spacecraft as part of K2 Campaign C14, with observations spanning  % CAT: is this a believable period?? seems slow
% no SDSS (?), 
UT 2017 June 25 to July 12, when \objname{} was at % 2457930.1821 to 2457946.6516, 16.470 days, 670 frames, 0.830 duty (?), 
heliocentric distance $4.196 < r_H < 4.271$~au, % via K2
% geocentric distance $3.581 < \Delta 3.874$~au, phase angle $12.359^\circ < \alpha < 13.626^\circ$,
and true anomaly angle $132.6^\circ < \nu < 134.1^\circ$. 
They produced a phase-folded lightcurve and found \objname{} had a rotation period of 22.493 days. % with a 0.18 magnitude amplitude.
We note this is roughly three orders of magnitude too slow to cause rotational disruption given the fiducial $\sim$2.2~hr spin-barrier threshold \citep{pravecFastSlowRotation2000}. % CAT would they have seen a tail? TODO

%%%%%%%%%%%%%%%%%%%%%%%%%%%%%%%%%%%%%%%%%%%%%%
\section{Thermodynamical Modeling}
\label{sec:thermo}

We conducted thermodynamical modeling to inform our activity mechanism analysis and assess prior posits that \objname{} should be inactive due to desiccation. Our model \citep{chandlerRecurrentActivityActive2021} follows \citet{hsiehMainbeltCometsPanSTARRS12015} in describing the temperature of an airless body on an orbit comparable to \objname{}. Here we utilize the measured geometric albedo ($A=0.045\pm0.041$; \citealt{mainzerNEOWISEDiametersAlbedos2019a}) to compute an estimated Bond albedo ($A_\mathrm{B}\approx \frac{2}{3} A \approx 0.03$) in our calculations, but consider a broad range of rotation scenarios despite the previously reported 22-day period.% TODO FINISH 5/18/2024 COC

\begin{figure*}
    \centering
    \includegraphics[width=0.75\linewidth]{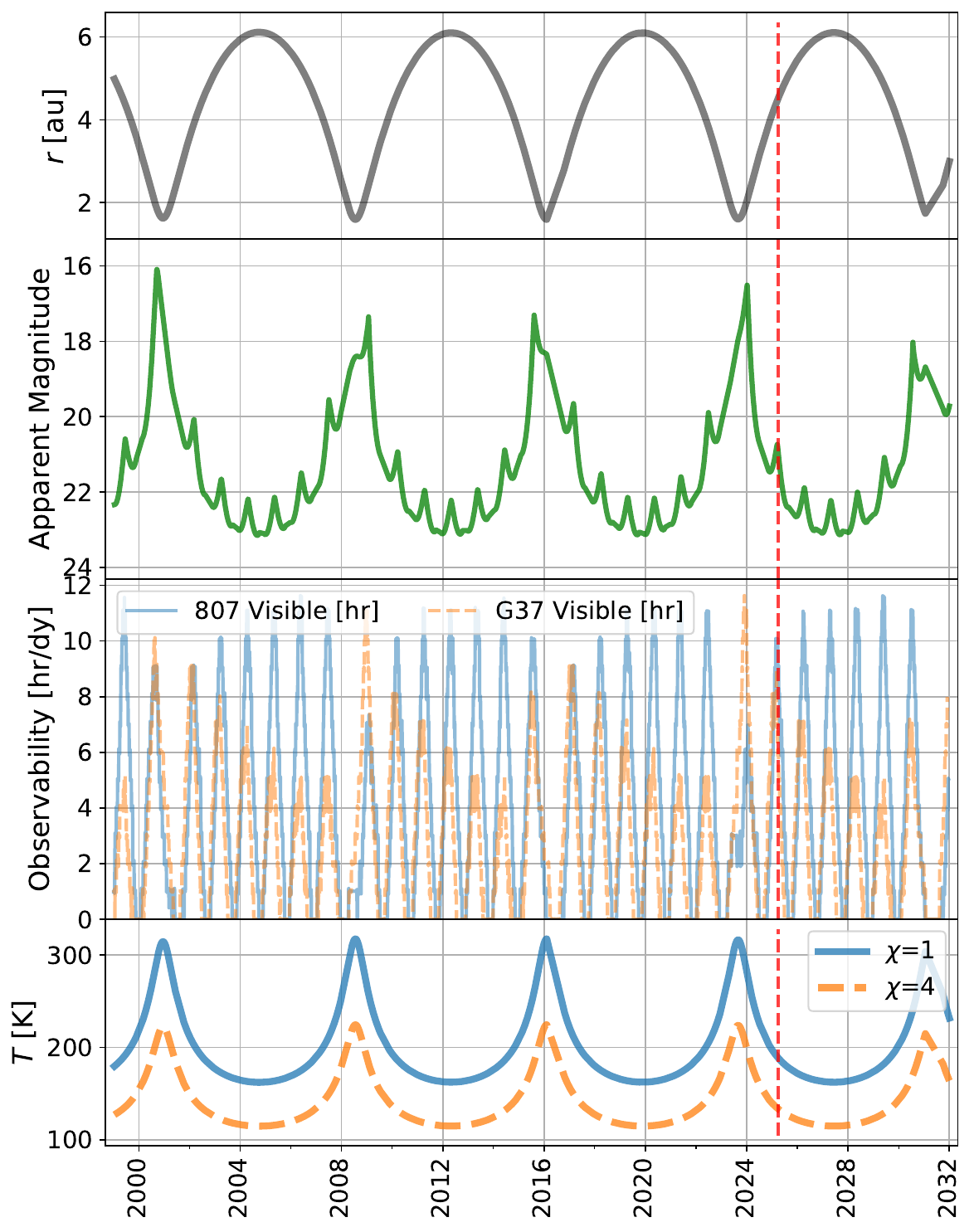}
    \caption{
    \objname{} metrics from UT 1999 January 1 to UT 2032 January 1, with UT 2025 April 1 indicated by a vertical dashed red line. % updated to longer timescale to show next observing window 10/1/2024 COC
    % \added{Above the heliocentric distance $r$ plot are indicated perihelion $q$ and aphelion $Q$ passages.}{} 
    Apparent $V$-band magnitude (via JPL Horizons). 
    ``Observability'' is the number of hours \objname{} was above the horizon for the indicated observatory site, in this case 807 (\acs{CTIO}, Chile) and G37 (Lowell Discovery Telescope near Happy Jack, Arizona), selected as representative examples for the southern and northern hemisphere sites, respectively. 
    % \added{\objname{} is again observable 2024 October.}{} 
    Solar oppositions and conjunctions can be identified here as corresponding coincident peaks and troughs in observability and apparent magnitude. 
    The modeled temperatures are bound by the thermophysical extremes: $\chi=1$ (solid line) the ``flat slab'' case, and $\chi=4$ (dashed line), the isothermal approximation. \objname{} has a reported period of $\sim$22 days, thus temperatures are more likely to be near the $\chi=1$ line.
    }
    \label{fig:observability}
\end{figure*}

We iteratively solve the equations of \cite{chandlerCometaryActivityDiscovered2020b} and input a range of temperatures that, in turn, return corresponding distances for a given equilibrium temperature. We then produce an inverse function that provides temperature as a function of heliocentric distance for a given set of parameters. As the albedo for \objname{} has been measured ($A\approx 0.05$; Table \ref{tab:params}), we hold this value constant. We compute temperatures corresponding to the thermophysical extremes of $\chi=1$ and $\chi=4$, as shown in Figure \ref{fig:observability}. While \cite{szaboRotationalPropertiesHilda2020} found the rotation period of \objname{} to be over 22 days, so $\chi=1$ (a ``flat slab'') is a more apt value, we include the full range of temperatures here in case the rotation period is revised in the future.

We combine our thermodynamical modeling results in an ``observability'' plot (Figure \ref{fig:observability}) that helps us find correlated events and identify potential observational biases. For example, activity is most likely detected when the object is brightest, i.e., at opposition and perihelion, with the caveat the object must have been visible at night from a given observatory. For sublimation-driven activity, observing an object at perihelion, where the temperature is warmest, helps maximize the likelihood of activity detection when the activity is caused by volatile sublimation. Figure \ref{fig:observability} shows the previous two perihelion passages (2015, 2008) took place during sub-optimal observing conditions (i.e., observable hours from either hemisphere, fainter apparent magnitude). However, when \objname{} was discovered, it was both bright and observable. Thus, the absence of observed activity suggests the object was quiescent during that apparition.

% \objname{} has passed perihelion four times since its discovery in 2000, but... 

Thermal fracture is another potential activity mechanism, whereby extreme temperatures or temperature swings cause a body to crack, releasing material in the process. This is thought to be a possible mechanism responsible for the Geminid meteor stream parent (3200)~Phaethon, which experiences temperatures from $\sim$195~K to over 800~K  \citep{ohtsukaSolarRadiationHeatingEffects2009,liRecurrentPerihelionActivity2013}. However, \objname{} reaches $\sim 310$~K at perihelion for the prescribed $\chi=1$, which, along with the relatively mild temperature swings ($\sim$100~K), suggests that thermal fracture is not responsible for the activity we observed. 

Throughout its present-day orbit, \objname{} remains warmer than 145~K, the gigayear-timescale surface water ice survival threshold for small solar system bodies \citep{schorghoferLifetimeIceMain2008,snodgrassMainBeltComets2017}. In Section \ref{sec:dynamics} we learn \objname{} is not stable for such a long period. Even if \objname{} was stable for such a period, we cannot rule out sublimation because it is still possible for subsurface ice as shallow as a few cm to survive \citep{schorghoferLifetimeIceMain2008,prialnikCanIceSurvive2009} which, in turn, can be excavated or exposed by another event, such as an impact, discussed further in Section \ref{sec:summary}. %\added{\cite{diazCollisionalActivationAsteroids2008}{} considered this scenario for over 200 asteroids on cometary orbits (\acsp{ACO}), including \objname{}, and found the timescale for an impact significant enough to excavate ices for sublimation to be on the order of once every 300,000 years. Thus it is unlikely that an impact is the cause of the activity we observed.}

%%%%%%%%%%%%%%%%%%%%%%%%%%%%%%%%%%%%%%%%%%%%%%%%%%%%%%%%%%%%%%%%%%%%%%%%%%%%%%%%%
\section{Dynamical Study}
\label{sec:dynamics}

Exploring the dynamical history of \objname{} can further our understanding of the body's potential composition, as well as inform our diagnosis of the underlying activity mechanism. Previously, \cite{gil-huttonCometCandidatesQuasiHilda2016} found \objname{} had an orbit consistent with a Centaur prior to -10.2 kyr, and was in the \ac{TNO} region prior to -27 kyr, at $a>300$~au \citep{garciamiganiActividadCometariaObjetos2019}. Moreover, they determined \objname{} could have been in an orbit with a perihelion distance under 1~au for $429 \pm 21$ years, which they describe as significant because objects with $q<1$ au for $>1$~kyr are likely to be dormant/desiccated, and that \objname{} is itself likely depleted. However, the \cite{gil-huttonCometCandidatesQuasiHilda2016} study was carried out nearly a decade ago, so we conducted a fresh dynamical investigation that incorporated the most up-to-date orbital elements known for \objname{}, as provided by JPL Horizons \citep{giorginiJPLsOnLineSolar1996}.
% asdf
% However, given the relatively recent arrival of \objname{} from its Centaur orbit [THAT IS GG DEPENDENT], the object would not have lost sufficient volatiles to be depleted. However, as GG point out, \objname{} spent a bunch of time near the Sun, which could have dessicated it, though they said over 1000 years and it wasn't there that long ($<500$~yr).

We carried out dynamical simulations with the \texttt{Rebound} package \citep{reinREBOUNDOpensourceMultipurpose2012} and the \texttt{Trace} \citep{luTRACECodeTimereversible2024a} integrator. We included the Sun and all of the planets (except for Mercury), and generated 500 orbital clones based on a Gaussian distribution of the orbital uncertainties (Table \ref{tab:orbitparameters}) as provided by JPL Horizons \citep{giorginiJPLsOnLineSolar1996}. We ran our simulations for $\pm 4$~Myr with a nominal 0.02~yr timestep and, crucially, variable timestep capability to properly account for close encounters.

Key results are provided in Figure \ref{fig:dynamics}, starting with general orbital configuration of \objname{} (Figure \ref{fig:dynamics}a) as viewed from above the plane of the solar system. Figure \ref{fig:dynamics}b shows \objname{} in the Jupiter co-rotating reference frame, a technique we employ to help distinguish between \acp{QHA} and \acp{JFC} \citep[e.g.,][]{chandlerMigratoryOutburstingQuasiHilda2022,oldroydRecurringActivityDiscovered2023}. In Figure \ref{fig:dynamics}c and b \objname{} can be seen periodically crossing the outer orbit of Mars. Notably, while \objname{} fully crosses the orbit of Jupiter (Figure \ref{fig:dynamics} panels a -- c), at present ($\pm$200 yr) the minor planet does not typically experience close encounters with the gas giant, where it normally stays exterior to one Jupiter Hill radius (Figure \ref{fig:dynamics}d). 

Figure \ref{fig:dynamics}e highlights \objname{}'s longitude of perihelion precession on a roughly 20~kyr cycle. This indicates \objname{} undergoes a period of close encounters with Jupiter that takes place roughly every 40~kyr. The next window of such encounters with Jupiter will be in about 35~kyr. 

Despite these periods of Jovian encounters, when we consider the dynamical history of \objname{} going back 4~Myr (Figure \ref{fig:dynamics}f), we find that the vast majority (99.2\%) of clones originate from the Oort cloud (the remaining 0.8\% were  \acp{JFC}). We also see that \objname{} most likely transitioned to the \ac{JFC} regime some 750~kyr ago, with 100\% of the clones belonging to the \acp{JFC} at -160~kyr. 

Our results differ substantially from those of \cite{gil-huttonCometCandidatesQuasiHilda2016} (especially their Figure 1, which shows the semimajor axis, perihelion, and aphelion evolution). We attribute these discrepancies to several factors, including a significant difference in orbital elements (e.g., the current value of $a$ is $>0.01$ au further out than their value). Moreover, their Bulirsch-Stoer \citep{bulirschNumericalTreatmentOrdinary1966} integrator was configured for a fixed one-day timestep, which may be inadequate for the close Jovian encounters encountered by \objname{}.

\objname{} has experienced on the order of 100,000 perihelion passages since it arrived in its present orbit (Figure \ref{fig:dynamics}f). This is far in excess of the more typical \ac{JFC} activity lifetime limit of $N_q=1600$ (equating to about 12,000 years) described by \cite{levisonKuiperBeltJupiterFamily1997}, but well within the prescribed dynamical lifetime for \acp{JFC}.

\renewcommand{\figsize}{0.48}
\begin{figure*}
    \centering
    \begin{tabular}{cc}
        \includegraphics[width=\figsize\linewidth]{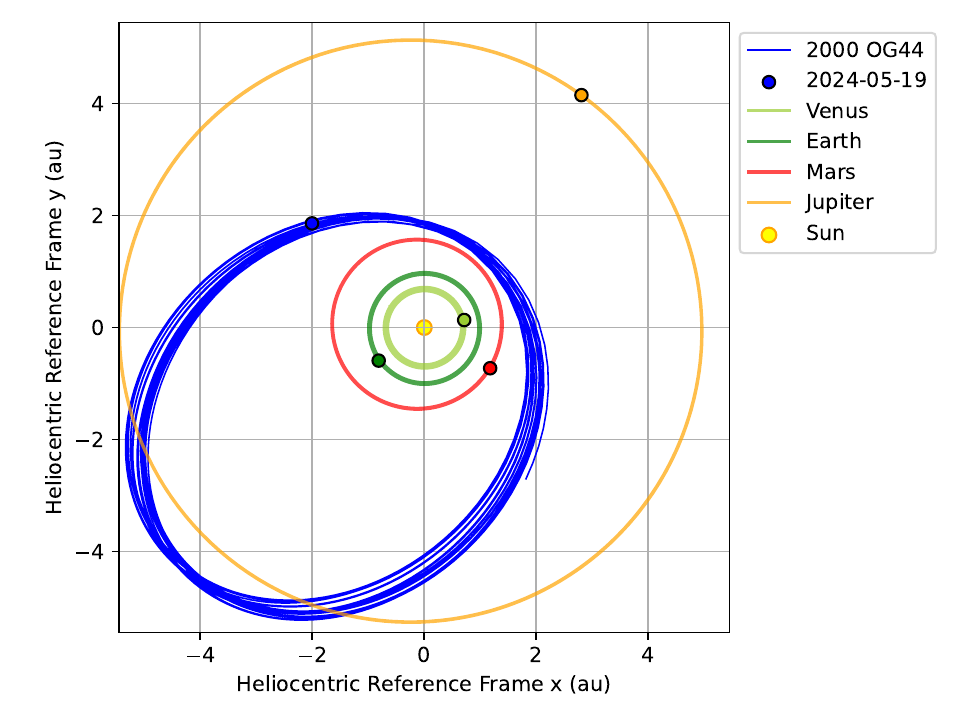}%{2000_OG44_orbit.png} 
        & 
                 \includegraphics[width=\figsize\linewidth]{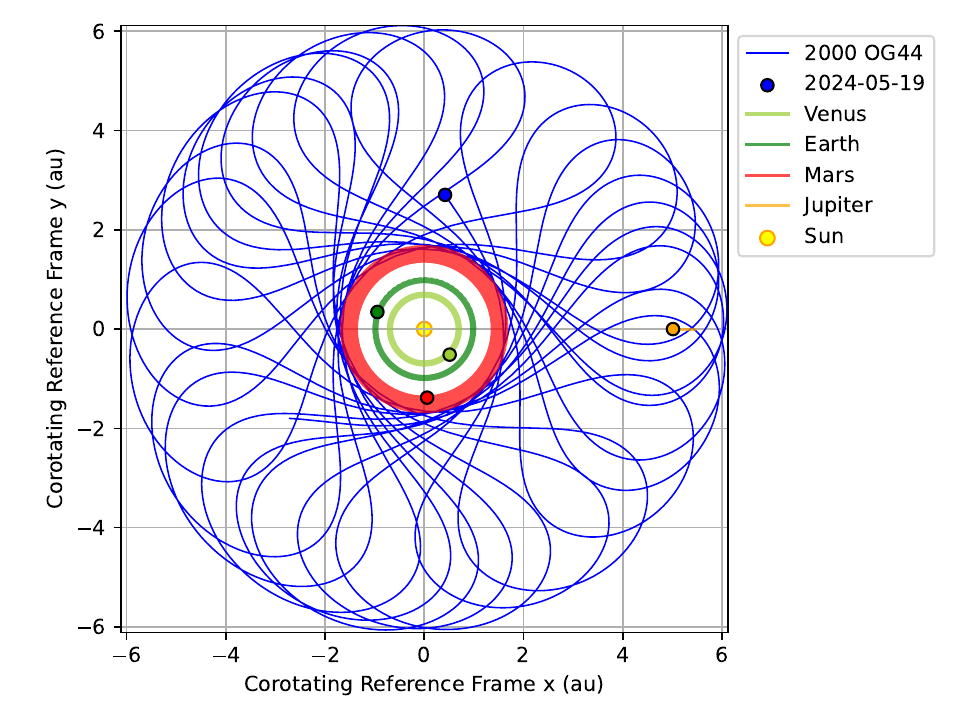}\\
                 (a) & (b)\\
        \\
         % {corotating_with_jupiter_2000_OG44.png}
         \includegraphics[width=\figsize\linewidth]{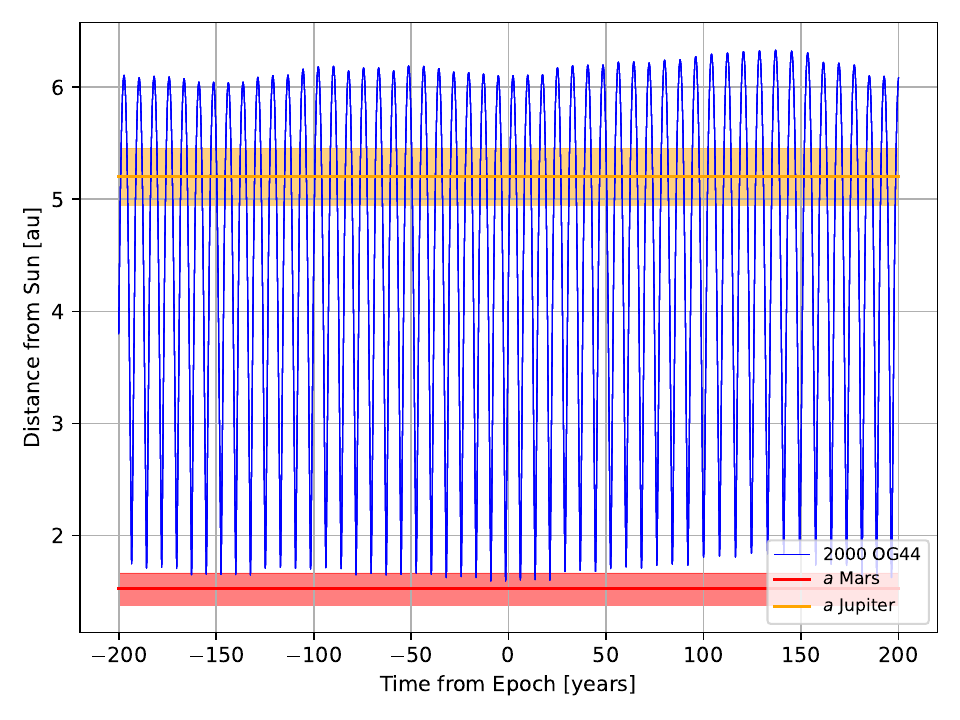} & \includegraphics[width=\figsize\linewidth]{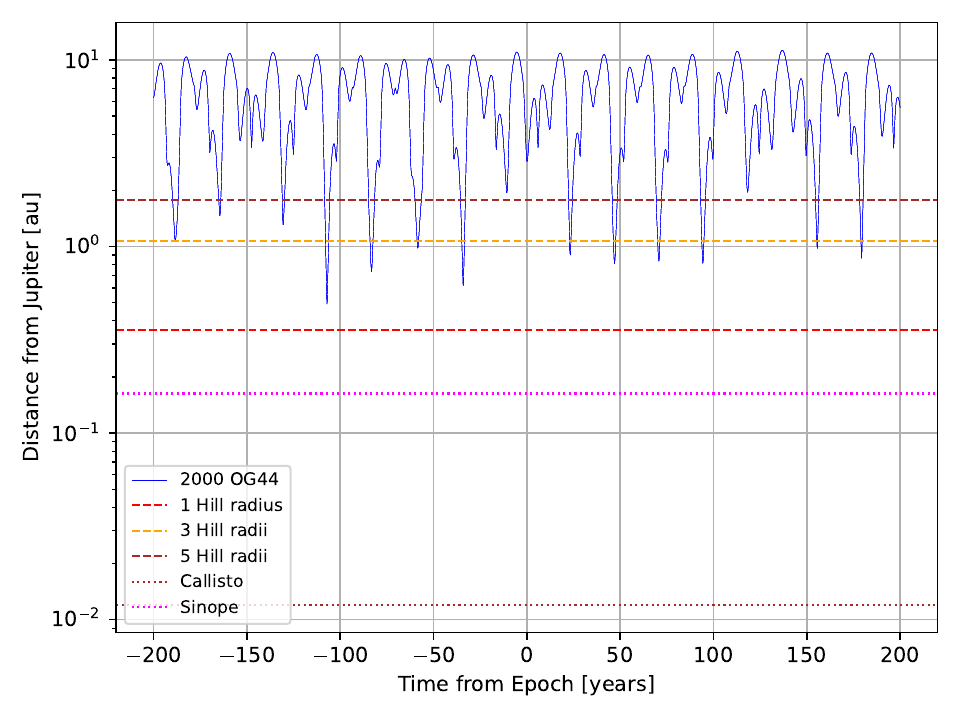}\\
         (c) & (d)\\
         \\
         \includegraphics[width=\figsize\linewidth]{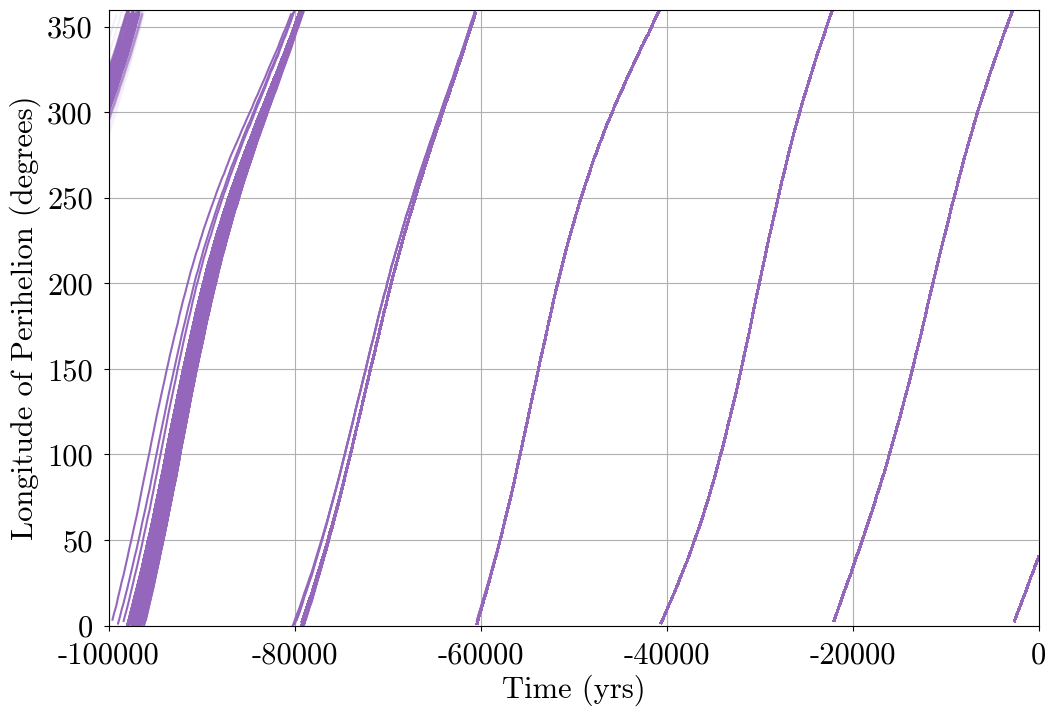} &

         \includegraphics[width=\figsize\linewidth]{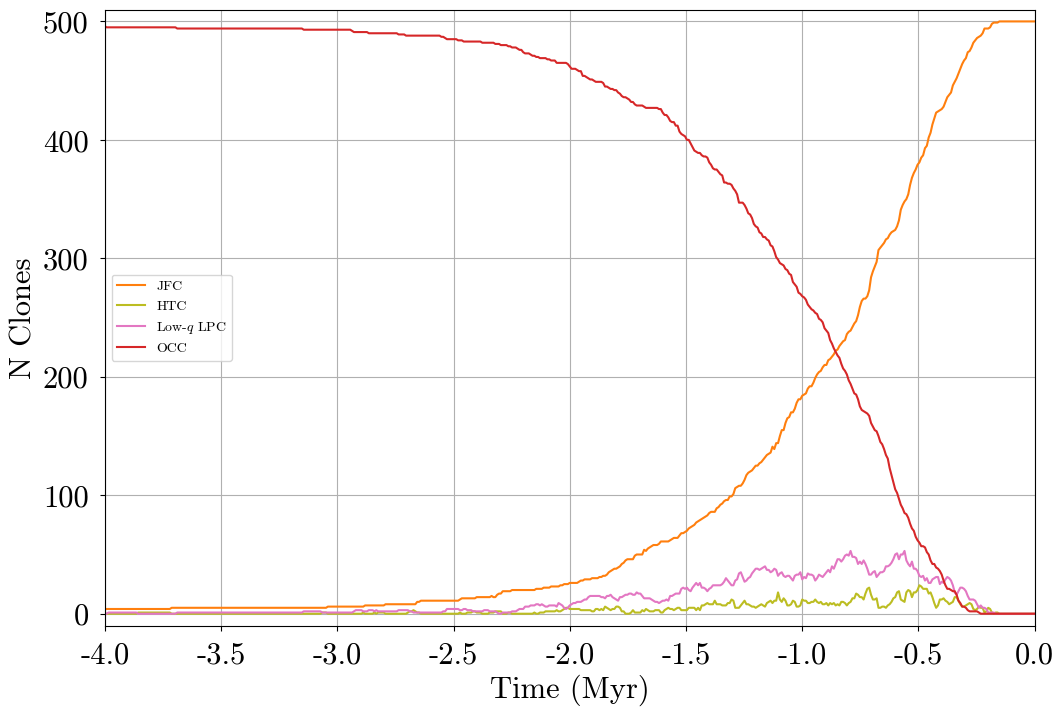} \\
         (e) & (f)\\
    \end{tabular}
    \caption{\objname{} dynamical modeling results. Panels (c) through (f) are comprised of 500 orbital clones spanning a Gaussian distribution of orbital element uncertainties. 
    \textbf{(a)} Solar system plot with \objname{}. 
    \textbf{(b)} Heliocentric distance of \objname{} and several planets, as seen from Jupiter's co-rotating reference frame. % timespan?
    \textbf{(c)} Heliocentric distance $r$. \objname{} crosses the orbit of Jupiter (orange horizontal line at $\sim$5~au) and Mars (red line at $\sim$1.5~au). Aphelion and perihelion distances for Jupiter and Mars are shown as shaded regions. 
    \textbf{(d)} Jupiter--\objname{} separation. 
    \textbf{(e)} Longitude of Perihelion precessing with a period of $\sim$20 kyr. 
    \textbf{(f)} Number of clones belonging to the named dynamical class; at 4 Myr ago, 99.2\% of clones originated in the Oort cloud, with the remaining 0.02\% belonging to the \acp{JFC}. % TODO clarify 3/20/2025 COC
    }
    \label{fig:dynamics}
\end{figure*}

%%%%%%%%%%%%%%%%%%%%%%%%%%%%%%%%%%%%%%%%%%%%%%%%%%%%%%%%%%%%%%%%%%%%
\section{Dynamical Classification}
\label{sec:class}

A commonly employed tool to distinguish between dynamical classes is the Tisserand Parameter with respect to Jupiter, which relates the influence of Jupiter over a body's orbit by 

\begin{equation}
    T_\mathrm{J} = \frac{a_\mathrm{J}}{a} + 2 \cos\left(i\right)\sqrt{\frac{a\left(1-e^2\right)}{a_\mathrm{J}}}, 
\end{equation}

\noindent where $a$ and $a_\mathrm{J}$ are the semi-major axes of \objname{} and Jupiter, respectively, $e$ is the orbital eccentricity of \objname{}, and $i$ is the orbital inclination of \objname{}. Cometary bodies typically have $T_\mathrm{J}<3$ \citep{levisonCometTaxonomy1996}, though the boundary is inexact \citep{hsiehPotentialJupiterFamilyComet2016}. Relevant here is the ambiguity between the \acp{JFC} and the quasi-Hilda objects --- bodies that have orbits similar to asteroid (153)~Hilda but are not themselves bound within the stable interior 3:2 \ac{MMR} with Jupiter. Dynamical modeling (Figure \ref{fig:dynamics}b) is needed to distinguish between the two, as we describe in  \cite{chandlerMigratoryOutburstingQuasiHilda2022} and \cite{oldroydCometlikeActivityDiscovered2023}. 
\objname{} has previously been reported as a \ac{QHC} candidate \citep{gil-huttonCometCandidatesQuasiHilda2016,correa-ottoPopulationCometCandidates2023} and a Hilda asteroid in 3:2 \ac{MMR} with Jupiter \cite{kankiewiczStabilityMostHazardous2006}. % In ``Stability of the Most Hazardous Mars-Crossers'' 
% \cite{kankiewiczStabilityMostHazardous2006} reported that \objname{} is in 3:2 \ac{MMR} with Jupiter. 

%

To explore the quasi-Hilda membership of \objname{} we consider here first the \cite{tothQuasiHildaSubgroupEcliptic2006} excitation parameter

% CAT: so why did everyone else get it wrong? Old parameters?; COC: as of now (8/9/2024) they seem to be within bounds.

\begin{equation}\label{eq:excitation}
    \mathcal{E} = \sqrt{e^2 + \sin^2(i)},
\end{equation}

\noindent where $e$ and $i$ are the orbital eccentricity and inclination. \cite{tothQuasiHildaSubgroupEcliptic2006} defines the quasi-Hildas as $2.90 < T_\mathrm{J} < 3.04$ and $0.18 < \mathcal{E} < 0.36$. The present-day values are $e=0.584$ and $i=7.396^\circ$, which results in an $\mathcal{E}=0.598$. For historic values we make use of the eccentricity and inclination reported in (1) the \ac{MPEC} 2000-S65 $e=0.5808864$, $i=7.32620^\circ$ \citep{hahnMPEC2000S6520002000}, and (2) \cite{gil-huttonCometCandidatesQuasiHilda2016} ($e=0.586$, $i=7.42^\circ$), which result in $\mathcal{E}=0.595$ and $\mathcal{E}= 0.600$, respectively. While both values fall outside the mean $\mathcal{E}=0.27\pm0.09$ quasi-Hilda excitement range described in \cite{tothQuasiHildaSubgroupEcliptic2006}, they point out notable outlier members, such as 77P/Longmore ($\mathcal{E}=0.547$) and 246P/NEAT ($\mathcal{E}=0.398$). Thus, while \objname{} does not have a typical \ac{QHC} excitation $\mathcal{E}$ metric, this property alone cannot rule out membership in the population, but \objname{} also has $T_\mathrm{J} = 2.735$, well outside the \cite{tothQuasiHildaSubgroupEcliptic2006} prescribed range.

A second method for identifying quasi-Hildas proposed by \cite{tothQuasiHildaSubgroupEcliptic2006} (also adopted by \cite{gil-huttonCometCandidatesQuasiHilda2016}) defines \acp{QHC} by their radial distance from a point in Lagrangian $(k,h)$ space, described by

\begin{equation}\label{eq:kh}
    k = e \cos\left(\bar{\omega}-\bar{\omega}_\mathrm{J}\right) \ \mathrm{and}\   
    h = e \sin\left(\bar{\omega}-\bar{\omega}_\mathrm{J}\right),
\end{equation}

\noindent where $e$ is the orbital eccentricity of the body, and $\bar{\omega}$ and $\bar{\omega_\mathrm{J}}$ are the longitude of perihelion for the object and Jupiter, respectively. We calculated the $(k,h)$ distance of \objname{} to be 0.518, well outside the \cite{tothQuasiHildaSubgroupEcliptic2006} prescribed range of an object being within a circle of radius 0.24 centered at $(k,h)=(0.075,0)$. Thus, by the $(k, h)$ metric, \objname{} does \added{not} qualify as a \ac{QHC}.

\objname{} has $T_\mathrm{J}=2.735$ (primarily due to its high eccentricity), which, along with our dynamical modeling, leads us to classify the object primarily as a \ac{JFC}. %We did not find \objname{} to be in 3:2 \ac{MMR}.
Moreover, with a perihelion distance $q=1.593$~au, interior to Mars' aphelion ($Q_\mathrm{M}=1.666$~au) but exterior to Mars' semi-major axis ($a_\mathrm{M}=1.530$~au), \objname{} is a Mars-crossing body of the ``outer grazer'' sub-type.
% WJO: where does the outer grazer definition come from?

%%%%%%%%%%%%%%%%%%%%%%%%%%%%%%%%%%%
\section{Summary and Discussion}\label{sec:summary}
% \acresetall{} % per WJO, done 4/2/2024 COC; % undone, per CAT, 5/18/2024 COC
% CAT: and it's a JFC and everyone else got it wrong
We observed \objname{}, an object that had previously been scrutinized for over 20 years in searches for cometary activity, yet none was found until this work. We found the object to be exhibiting a distinct, thin comet tail in the coincident anti-solar and anti-motion directions (as projected on the sky) in observations we carried out with  \ac{ARCTIC} on the 3.5~m \ac{ARC} at \ac{APO} on UT 2023 July 24 and 26 (Section \ref{sec:observations}). At that time, \objname{} was at a heliocentric distance $r_H=1.649$~au, inbound (true anomaly angle $\nu=338^\circ$) to its perihelion passage ($q=1.604$~au). 

Our dynamical modeling (Section \ref{sec:dynamics}) indicates \objname{} originated in the Oort cloud and transitioned to a \ac{JFC} orbit at least 100 kyr ago. We also saw that \objname{} regularly enters periods of close encounters with Jupiter, with the next cycle due in approximately 35 kyr. % years that prevents us from determining the fate of \objname{} beyond that encounter.
% asdf
\objname{} has experienced on the order of 100,000 perihelion passages since it arrived in its present orbit (Figure \ref{fig:dynamics}f), far in excess of the more typical \ac{JFC} limit of $N_q=1600$ (equating to about 12,000 years; \citealt{levisonKuiperBeltJupiterFamily1997}). 

% TODO redo everything with the new Oort cloud story! 5/18/2024 COC
We carried out thermodynamical modeling (Section \ref{sec:thermo}), informed by past ground- and space-based measurements found in the literature, and estimate that \objname{} experiences temperatures ranging from approximately 100~K to 300~K. Considering the observed 22-day rotation period \citep{szaboRotationalPropertiesHilda2020}, at all times \objname{} is at an equilibrium temperature above the 145~K surface water ice gigayear survival threshold \citep{schorghoferLifetimeIceMain2008,snodgrassMainBeltComets2017}. However, sublimation is still possible because 
(1) it is probable that \objname{} arrived much more recently than this timespan (Figure \ref{fig:dynamics}f) due to its dynamical behavior, 
(2) subsurface water ice could still be present, even on longer timescales \citep{schorghoferLifetimeIceMain2008,prialnikCanIceSurvive2009}, and 
(3) if \objname{} has a faster rotation rate, the object spends a significant fraction of its time in a region below 145~K (Figure \ref{fig:observability}), in which case water ice would survive. 
Regardless of the rotation rate, the temperature range and maximum temperature are unlikely to cause thermal fracture akin to that experienced by active bodies such as (3200) Phaethon \citep{ohtsukaSolarRadiationHeatingEffects2009,liRecurrentPerihelionActivity2013}. Moreover, %\added{\objname{} has a prolonged rotation period ($>22$ days; \citealt{szaboRotationalPropertiesHilda2020}), so we do not consider}{} 
rotational disruption is highly unlikely to be a viable activity mechanism given its reported long rotation period.

Another source of cometary activity is impact events, suggested in cases such as e.g., (596) Scheila \citep{moreno596ScheilaOutburst2011} and, notably, definitively caused in the case of Double Asteroid Redirection Test (DART) target Dimorphos, the moon of (65803) Didymos \citep{dalySuccessfulKineticImpact2023}. It is possible, but very unlikely (once every $\sim$300,000 years for a 100~m impactor; \citealt{diazCollisionalActivationAsteroids2008}) that \objname{} experienced an impact event that caused the activity we observe. We find it more likely that volatile sublimation is the primary activity driver given the apparent onset of activity near its perihelion passage, and additional observations of activity in the future would lend further support to this diagnosis. Moreover, \objname{} has been classified as a D-type object (Table \ref{tab:params}; \citealt{licandroVisibleNearinfraredSpectra2018}), which are thought to be primordial bodies that migrated inwards from the Kuiper belt \citep{levisonContaminationAsteroidBelt2009}, and thus would likely contain ices.%as predicted by the Nice model. %  [CITE].  % TODO REWORD 11/10/2023 COC

\objname{} has previously been dynamically classified as a comet and quasi-Hilda object, and qualifies by the definitions of \cite{tothQuasiHildaSubgroupEcliptic2006} and \cite{gil-huttonCometCandidatesQuasiHilda2016}. Our analysis (Section \ref{sec:class}) suggests that the object is, at present, on an orbit consistent with a member of the \acp{JFC}.

We encourage follow-up observations to help characterize the activity circumstances and morphology, noting that, if cometary activity is detectable again during this apparition, this will be the last opportunity until the object approaches perihelion again in late-2030. We observed \objname{} active during the inbound portion of its orbit, so we anticipate it will again be active again at $\nu \gtrsim 330^\circ$.

%%%%%%%%%%%%%%%%%%%%%%%%%%%%%%%%%%%%%%%%%%%%%%%%%%%%%%%%%%%%%%%%

\section*{Acknowledgements}

The authors wish to thank the anonymous referee, whose comprehensive feedback greatly enhanced the quality of this work.

Many thanks to Arthur and Jeanie Chandler for their ongoing support. % 3/21/2024 COC
We thank Dr.\ Mark Jesus Mendoza Magbanua (University of California San Francisco) and Nima Sedaghat (University of Washington) for their continued encouragement. 
We also thank Andrew Connolly (eScience Institute, LINCC Frameworks, University of Washington) and the other members of the University of Washington/DiRAC Institute Solar System Group, especially: 
Pedro Bernardinelli, 
Dino Bektešević, 
Max Frissell, 
Sarah Greenstreet, 
Ari Heinze, 
Mario Jurić, 
Bryce Kalmbach, 
Jake Kurlander, 
Chester Li, 
Naomi Morato, 
Joachim Moeyens, 
Steven Stetzler, 
and 
Yasin Arafi Chowdhury.
% TODO add Yasin 9/10/2024 COC

We thank Ben Williams (University of Washington) and the \ac{APO} team, especially Candace Gray, Amanda Townsend, Russet McMillan, and Jack Dembicky. We thank David Wang (University of Washington), Eric Bellm (University of Washington), and Tobin Wainer (University of Washington) for their generous telescope time contributions. 

C.O.C. and C.A.T. %, H.H.H., and C.A.T.\ 
acknowledge support from NASA grant 80NSSC19K0869. 
W.J.O. and C.A.T.\ acknowledge support from NASA grant 80NSSC21K0114. C.A.T. and C.O.C. acknowledge support from NASA grant 80NSSC24K1323 and NSF award 2408827. 
W.A.B. acknowledges support from NSF award 1950901. 
D.E.V. acknowledges that this material is based upon work supported by the National Science Foundation under Grant No. (2307569). This research award is partially funded by a generous gift of Charles Simonyi to the NSF Division of Astronomical Sciences. The award is made in recognition of significant contributions to Rubin Observatory’s Legacy Survey of Space and Time. %(NAU REU program in astronomy and planetary science). % 7/12/2022 COC -- per DET, for WAB?
This research received support through Schmidt Sciences. % 3/21/2024 COC new text
Chandler and Vavilov acknowledge support from the DIRAC Institute in the Department of Astronomy at the University of Washington. The DiRAC Institute is supported through generous gifts from the Charles and Lisa Simonyi Fund for Arts and Sciences, and the Washington Research Foundation. % added LINCC, DiRAC 7/3/2023 COC; if one has to go: DiRAC. % Commenting DiRAC 7/5/2023 COC
 Computational analyses were run on Northern Arizona University's Monsoon computing cluster, funded by Arizona's Technology and Research Initiative Fund. % 

Based on observations obtained with the \acl{APO} 3.5-meter telescope, which is owned and operated by the \acl{ARC}. Observations made use of \acf{ARCTIC} imager \citep{huehnerhoffAstrophysicalResearchConsortium2016a}. \acs{ARCTIC} data reduction made use of the \texttt{acronym} software package \citep{l.weisenburgerAcronymAutomaticReduction2017}. % Apache Point

The \acs{VATT} referenced herein refers to the Vatican Observatory’s Alice P. Lennon Telescope and Thomas J. Bannan Astrophysics Facility. We are grateful to the Vatican Observatory for the generous time allocations. A special thanks to Vatican Observatory Director Br. Guy Consolmagno, S.J., Rev.~Pavel Gabor, S.J., Rev. Richard P. Boyle, S.J., Gary Gray, Chris Johnson, and Michael Franz. 

World Coordinate System corrections facilitated by \textit{Astrometry.net} \citep{langAstrometryNetBlindAstrometric2010}.  % 7/5/2023 COC: technically not for this Note
This research has made use of 
 NASA's Astrophysics Data System, % skipping 7/5/2023 COC
 data and/or services provided by the International Astronomical Union's Minor Planet Center, 
% , and % 7/5/2023 COC: out due to word limit
SAOImageDS9, developed by Smithsonian Astrophysical Observatory \citep{joyeNewFeaturesSAOImage2006}, 
and the Lowell Observatory Asteroid Orbit Database \textit{astorbDB} \citep{bowellPublicDomainAsteroid1994,moskovitzAstorbDatabaseLowell2021}. 

This work made use of 
the \acs{CADC} Solar System Object Information Search \citep{gwynSSOSMovingObjectImage2012} 
and the Asteroid Lightcurve Database \citep{warnerAsteroidLightcurveDatabase2009}.

\vspace{5mm}
\facilities{
ARC:3.5m (ARCTIC), 
LDT (LMI), 
VATT (VATT4K)
}

\software{
        {\tt astropy} \citep{robitailleAstropyCommunityPython2013}, 
        {\tt astrometry.net} \citep{langAstrometryNetBlindAstrometric2010}, % already in acks 1/31/2023 COC
        {\tt JPL Horizons} \citep{giorginiJPLsOnLineSolar1996}, % already in acks, text 1/31/2023 COC
        {\tt Matplotlib} \citep{hunterMatplotlib2DGraphics2007}, % 7/5/2023 COC: I guess not for this. length limits
        {\tt NumPy} \citep{harrisArrayProgrammingNumPy2020}, % 7/5/2023 COC: I guess not for this. length limits
        {\tt pandas} \citep{rebackPandasdevPandasPandas2022}, % limiting to 1 citation, removing older mckinneyDataStructuresStatistical2010 1/31/2023 COC% 7/5/2023 COC: I guess not for this. length limits
        {\tt REBOUND} \citep{reinREBOUNDOpensourceMultipurpose2012},
        {\tt SAOImageDS9} \citep{joyeNewFeaturesSAOImage2006}, % 7/5/2023 COC: I guess not for this. length limits
        {\tt SciPy} \citep{virtanenSciPy10Fundamental2020}, % 7/5/2023 COC: I guess not for this. length limits
         {\tt Siril} \citep{richardSirilAdvancedTool2024},
       {\tt SkyBot} \citep{berthierSkyBoTNewVO2006} % cited in acks already
          }

%%%%%%%%%%%%%%%%%%%%%%%%%%%%%%%%%%%%%%%%%%%%%%%%%%%

\appendix

\section{Orbital Elements}\label{appendix:orbitalelements}

\begin{table}[h]
\caption{\objname{} Orbital Parameters}
\label{tab:orbitparameters} % NOTE: MUST COME AFTER CAPTION COMMAND! 11/23/2024 COC
\begin{tabular}{lllcrh}
Symbol & Parameter                          & Value                & Uncertainty & Units  & Source(s) \\
\hline
$a$      & Semi-major Axis                    & 3.85786413           & 6.95E-09    & au     & JPL       \\
$e$      & Eccentricity                       & 0.584249229          & 1.44E-09    &        & JPL       \\
$i$      & Inclination                        & 7.396408697          & 1.12E-07    & deg    & JPL       \\
$q$      & Perihelion Distance                & 1.603909987          & 4.21E-09    & au     & JPL       \\
$Q$      & Aphelion Distance                  & 6.111818273          & 1.10E-08    & au     & JPL       \\
$P$      & Period                             & 7.577546229          & 2.05E-08    & yr     & JPL       \\
$T_\mathrm{J}$     & Tisserand Parameter w.r.t. Jupiter & 2.735                &             &        & JPL       \\
node   & Longitude of the Ascending Node    & 301.3503443          & 5.06E-07    & deg    & JPL       \\
peri   & Argument of Perihelion             & 98.87864396          & 6.58E-07    & deg    & JPL       \\
$t_\mathrm{p}$     & Time of Perihleion Passage (TDB)         & 2023-Aug-30.57824219 & 7.64E-06    & dy    & JPL       \\
% $n$      & Mean Motion                        & 0.130071959          & 3.51E-10    & deg/dy & JPL   
& Epoch (UT TDB) & 2023-Sep-13.0 & & & JPL\\
& Epoch (JD TDB) & 2460200.5 & & dy & JPL\\
\end{tabular}

\raggedright
\footnotesize

Orbital elements retrieved UT 2023 November 4 from the JPL Small-Body Database \citep{giorginiJPLsOnLineSolar1996}
\end{table}

\objname{} orbital elements are provided in Table \ref{tab:orbitparameters}.

%%%%%%%%
\section{Catalog of Measurements}

\begin{table*}[h]
\centering
\caption{\objname{} Measurements from Literature}
\label{tab:params}
\begin{tabular}{clllccccc}
Symbol & Parameter               & Value      & Uncertainty & Units    & Date       & $r_\mathrm{H}$ [au]    & $f$ [$^\circ$]    & Source \\
\hline
$J-H$    & J-H Color               & 0.420      & 0.058    &          & 2000-08-25 & 1.964 & 302   & 1      \\
$J-H$    & J-H Color               & 0.447      & 0.058    &          & 2000-08-25 & 1.964 & 302   & 1      \\
$H-K$    & H-K Color               & 0.103      & 0.085    &          & 2000-08-25 & 1.964 & 302   & 1      \\
$H-K$    & H-K Color               & 0.175      & 0.085    &          & 2000-08-25 & 1.964 & 302   & 1      \\
$H$      & Absolute Magnitude      & 14.2       &             &          & 2000-10-09 & 1.757 & 323   & 2      \\
$G$      & Slope                   & 0.30700    &             &          & 2000-10-09 & 1.757 & 323   & 2      \\
       & Spectral class          & D          &             &          & 2000-10-09 & 1.757 & 323   & 2      \\
$D$      & Diameter                & 5.6        & 0.6         & km       & 2000-11-08 & 1.665 & 339   & 3      \\
$A_R$   & R-band Geometric Albedo & 0.111      & 0.024       &          & 2000-11-08 & 1.665 & 339   & 3      \\
$\eta$    & Beaming Parameter       & 0.4 to 0.6 &             &          & 2000-11-08 & 1.665 & 339   & 3      \\
       & Visible Slope           & 8.88       &             & \%/1000Å & 2014-07-22 & 4.476 & 222   & 4      \\
       & Spectral class          & D          &             &          & 2014-07-22 & 4.476 & 222   & 4      \\
$H$      & Absolute Magnitude      & 14.5       &             &          & 2015-10-29 & 1.901 & 304   & 5      \\
$G$      & Slope                   & 0.15       &             &          & 2015-10-29 & 1.901 & 304   & 5      \\
$D$      & Diameter                & 8.49       & 2.249       & km       & 2015-10-29 & 1.901 & 304   & 5      \\
       & Geometric Albedo        & 0.045      & 0.041       &          & 2015-10-29 & 1.901 & 304   & 5      \\
       & IR Albedo               & 0.068      & 0.107       &          & 2015-10-29 & 1.901 & 304   & 5      \\
$\eta$    & Beaming Parameter       & 0.95       & 0.2         &          & 2015-10-29 & 1.901 & 304   & 5     \\
p      & Rotation period         & 22.493     &             & dy       & 2017-06-25 & 4.196 & 133 & 6      \\
$\cdots$      & $\cdots$                       & $\cdots$          &             & $\cdots$        & 2017-07-12 & 4.271 & 134 & 6      \\
\end{tabular}

\raggedright
\footnotesize
Heliocentric distance $r_\mathrm{H}$ and true anomaly angle $\nu$ were retrieved UT 2024 May 18 from JPL Horizons \citep{giorginiJPLsOnLineSolar1996}. 
The rotation period was measured between the two dates provided. 
Sources: 
1. \cite{sykes2MASSAsteroidComet2000,sykes2MASSAsteroidComet2010} using \ac{2MASS}. 
2. \cite{binzelCompositionalDistributionsEvolutionary2019} \ac{MITHNEOS} with the 3~m \ac{IRTF} (Manua Kea, Hawaii). 
3. \cite{fernandezAlbedosAsteroidsCometLike2005} with Keck and the Hawaii 88'' (Manua Kea, Hawaii). 
4. \cite{licandroVisibleNearinfraredSpectra2018} with the 2.5~m \ac{INT} at the El Roque de los Muchachos Observatory in La Palma, Spain. 
5. \cite{mainzerNEOWISEDiametersAlbedos2019a} via \ac{NEOWISE}. 
6. \cite{szaboRotationalPropertiesHilda2020} with K2. 
\end{table*}

Table \ref{tab:params} lists measurements collected from the literature referenced during our investigation.

\clearpage
\bibliography{zotero}{}
\bibliographystyle{aasjournalv7} % to v7 4/8/2025 COC

%% This command is needed to show the entire author+affiliation list when the collaboration and author truncation commands are used. It has to go at the end of the manuscript.
%\allauthors

%% Include this line if you are using the \added, \added, \deleted commands to see a summary list of all changes at the end of the article.
%\listofchanges
\input{Acronyms.tex}
\end{document}

%% file: Acronyms.tex
% \section*{Acronyms}
% \label{sec:acronyms}
% \begin{acronym}
\begin{acronym}[JSONP]\itemsep0pt % 9/11/2023 COC -- single space
\acro{2MASS}{Two Micron All Sky Survey}
\acro{AA}{active asteroid}
\acro{ACO}{asteroid on a cometary orbit}
\acro{AI}{artificial intelligence}
\acro{API}{Application Programming Interface}
\acro{APT}{Aperture Photometry Tool}
\acro{ARC}{Astrophysical Research Consortium}
\acro{ARCTIC}{Astrophysical Research Consortium Telescope Imaging Camera}
\acro{APO}{Apache Point Observatory}
\acro{ARO}{Atmospheric Research Observatory}
\acro{AstOrb}{Asteroid Orbital Elements Database}
\acro{ASU}{Arizona Statue University}
\acro{AURA}{Association of Universities for Research in Astronomy}
\acro{BASS}{Beijing-Arizona Sky Survey}
\acro{BLT}{Barry Lutz Telescope}
\acro{CADC}{Canadian Astronomy Data Centre}
\acro{CASU}{Cambridge Astronomy Survey Unit}
\acro{CATCH}{Comet Asteroid Telescopic Catalog Hub}
\acro{CBAT}{Central Bureau for Astronomical Telegrams}
\acro{CBET}{Central Bureau for Electronic Telegrams}
\acro{CCD}{charge-coupled device}
\acro{CEA}{Commissariat a l'Energes Atomique}
\acro{CFHT}{Canada France Hawaii Telescope}
\acro{CFITSIO}{C Flexible Image Transport System Input Output}
\acro{CNEOS}{Center for Near Earth Object Studies} % 9/11/2023 COC
\acro{CNRS}{Centre National de la Recherche Scientifique}
\acro{CPU}{Central Processing Unit}
\acro{CTIO}{Cerro Tololo Inter-American Observatory}
\acro{DAPNIA}{Département d'Astrophysique, de physique des Particules, de physique Nucléaire et de l'Instrumentation Associée}
\acro{DART}{Double Asteroid Redirection Test}
\acro{DDT}{Director's Discretionary Time}
\acro{DECaLS}{Dark Energy Camera Legacy Survey}
\acro{DECam}{Dark Energy Camera}
\acro{DES}{Dark Energy Survey}
\acro{DESI}{Dark Energy Spectroscopic Instrument}
\acro{DCT}{Discovery Channel Telescope}
\acro{DOE}{Department of Energy}
\acro{DR}{Data Release}
\acro{DS9}{Deep Space Nine}
\acro{ESO}{European Space Organization}
\acro{ETC}{exposure time calculator}
\acro{ETH}{Eidgenössische Technische Hochschule}
\acro{FAQ}{frequently asked questions}
\acro{FITS}{Flexible Image Transport System}
\acro{FOV}{field of view}
\acro{GEODSS}{Ground-Based Electro-Optical Deep Space Surveillance}
\acro{GIF}{Graphic Interchange Format}
\acro{GMOS}{Gemini Multi-Object Spectrograph}
\acro{GPU}{Graphics Processing Unit}
\acro{GRFP}{Graduate Research Fellowship Program}
\acro{HARVEST}{Hunting for Activity in Repositories with Vetting-Enhanced Search Techniques}
\acro{HSC}{Hyper Suprime-Cam}
\acro{IAU}{International Astronomical Union}
\acro{IMACS}{Inamori-Magellan Areal Camera and Spectrograph}
\acro{IMB}{inner Main-belt}
\acro{IMCCE}{Institut de Mécanique Céleste et de Calcul des Éphémérides}
\acro{INAF}{Istituto Nazionale di Astrofisica}
\acro{INT}{Isaac Newton Telescopes}
\acro{IP}{Internet Protocol}
\acro{IRSA}{Infrared Science Archive}
\acro{IRTF}{Infrared Telescope Facility}
\acro{ITC}{integration time calculator}
\acro{JAXA}{Japan Aerospace Exploration Agency}
\acro{JD}{Julian Date}
\acro{JFC}{Jupiter-family comet}
\acro{JPL}{Jet Propulsion Laboratory}
\acro{KBO}{Kuiper Belt object}
\acro{KOA}{Keck Observatory Archive}
\acro{KPNO}{Kitt Peak National Observatory}
\acro{LBC}{Large Binocular Camera}
\acro{LCO}{Las Campanas Observatory}
\acro{LBCB}{Large Binocular Camera Blue}
\acro{LBCR}{Large Binocular Camera Red}
\acro{LBT}{Large Binocular Telescope}
\acro{LDT}{Lowell Discovery Telescope}
\acro{LINCC}{LSST Interdisciplinary Network for Collaboration and Computing}
\acro{LINEAR}{Lincoln Near-Earth Asteroid Research}
\acro{LMI}{Large Monolithic Imager}
\acro{LONEOS}{Lowell Observatory Near-Earth-Object Search}
\acro{LSST}{Legacy Survey of Space and Time}
\acro{MBC}{Main-belt comet}
\acro{MGIO}{Mount Graham International Observatory}
\acro{MITHNEOS}{MIT-Hawaii Near-Earth Object Spectroscopic Survey}
\acro{ML}{machine learning}
\acro{MMB}{middle Main-belt}
\acro{MMR}{mean-motion resonance}
\acro{MOST}{Moving Object Search Tool}
\acro{MPEC}{Minor Planet Electronic Circular} % 9/10/2024 COC
\acro{MzLS}{Mayall z-band Legacy Survey}
\acro{MPC}{Minor Planet Center}
\acro{NAU}{Northern Arizona University}
\acro{NEA}{near-Earth asteroid}
\acro{NEAT}{Near-Earth Asteroid Tracking}
\acro{NEATM}{Near Earth Asteroid Thermal Model}
\acro{NEO}{near-Earth object}
\acro{NEOWISE}{Near-Earth Object Wide-field Infrared Survey Explorer}
\acro{NIHTS}{Near-Infrared High-Throughput Spectrograph}
\acro{NOAO}{National Optical Astronomy Observatory}
\acro{NOIRLab}{National Optical and Infrared Laboratory}
\acro{NRC}{National Research Council}
\acro{OMB}{outer Main-belt}
\acro{OSIRIS-REx}{Origins, Spectral Interpretation, Resource Identification, Security, Regolith Explorer}
\acro{NSF}{National Science Foundation}
\acro{PANSTARRS}{Panoramic Survey Telescope and Rapid Response System.}
\acro{PI}{Principal Investigator}
\acro{PNG}{Portable Network Graphics}
\acro{PSI}{Planetary Science Institute}
\acro{PSF}{point spread function}
\acro{PTF}{Palomar Transient Factory}
\acro{QH}{Quasi-Hilda}
\acro{QHA}{Quasi-Hilda Asteroid}
\acro{QHC}{Quasi-Hilda Comet}
\acro{QHO}{Quasi-Hilda Object}
\acro{RA}{Right Ascension}
\acro{REU}{Research Experiences for Undergraduates}
\acro{RNAAS}{Research Notes of the American Astronomical Society}
\acro{SAFARI}{Searching Asteroids For Activity Revealing Indicators}
\acro{SDSS}{Sloan Digital Sky Survey}
\acro{SMOKA}{Subaru Mitaka Okayama Kiso Archive}
\acro{SAO}{Smithsonian Astrophysical Observatory}
\acro{SBDB}{Small Body Database}
\acro{SDSS DR-9}{Sloan Digital Sky Survey Data Release Nine}
\acro{SLAC}{Stanford Linear Accelerator Center}
\acro{SOAR}{Southern Astrophysical Research Telescope}
\acro{SNR}{signal to noise ratio}
\acro{SSOIS}{Solar System Object Information Search}
\acro{SQL}{Structured Query Language}
\acro{SUP}{Suprime Cam}
\acro{SWRI}{Southwestern Research Institute}
\acro{TAP}{Telescope Access Program}
\acro{TNO}{Trans-Neptunian object}
\acro{UA}{University of Arizona}
\acro{UCSC}{University of California Santa Cruz}
\acro{UCSF}{University of California San Francisco}
\acro{VATT}{Vatican Advanced Technology Telescope}
\acro{VIA}{Virtual Institute of Astrophysics}
\acro{VIRCam}{VISTA InfraRed Camera}
\acro{VISTA}{Visible and Infrared Survey Telescope for Astronomy}
\acro{VLT}{Very Large Telescope}
\acro{VST}{Very Large Telescope (VLT) Survey Telescope}
\acro{WFC}{Wide Field Camera}
\acro{WIRCam}{Wide-field Infrared Camera}
\acro{WISE}{Wide-field Infrared Survey Explorer}
\acro{WCS}{World Coordinate System}
\acro{YORP}{Yarkovsky--O'Keefe--Radzievskii--Paddack}
\acro{ZTF}{Zwicky Transient Facility}
\end{acronym}